\documentclass[journal=ancac3,manuscript=article]{achemso}

\usepackage{chemformula} % Formula subscripts using \ch{}
\usepackage[utf8]{inputenc} %For the ``Umlaute'' auf Windows PCs
\usepackage[T1]{fontenc} % Use modern font encodings
\usepackage{graphicx} %used with \includegraphics{}
\usepackage{wrapfig}
\usepackage{lineno}

\author{Huanyao Cun}
\email{hycun1@physik.uzh.ch}
\affiliation{Physik-Institut, Universit\"at Z\"urich, 8057 Z\"urich, Switzerland}
\altaffiliation{Contributed equally to this work}
\email{hycun1@physik.uzh.ch}

\author{Zichun Miao}
\affiliation{Department of Physics, Hong Kong University of Science and Technology, Kowloon, Hong~Kong, China}
\altaffiliation{Contributed equally to this work}

\author{Adrian Hemmi}
\affiliation{Physik-Institut, Universit\"at Z\"urich, 8057 Z\"urich, Switzerland}

\author{Marcella Iannuzzi}
\affiliation{Department of Chemistry, University of Zurich, 8057 Z\"urich, Switzerland}

\author{J\"urg Osterwalder}
\affiliation{Physik-Institut, Universit\"at Z\"urich, 8057 Z\"urich, Switzerland}

\author{Michael S. Altman}
\affiliation{Department of Physics, Hong Kong University of Science and Technology, Kowloon, Hong~Kong, China}

\author{Thomas Greber}
\affiliation{Physik-Institut, Universit\"at Z\"urich, 8057 Z\"urich, Switzerland}

\title{High-quality hexagonal boron nitride from 2D distillation}

\keywords{$h$-BN , 2D materials transfer, moir\'e, heterogeneous catalysis, 2D distillation}

\begin{document}


%\begin{tocentry}
%\includegraphics[scale=0.39]{TOC.pdf}
 % \label{fig:TOC}
%\end{tocentry}

%\begin{TOC}

%\begin{wrapfigure}{L}{0.52\textwidth}
%\includegraphics[width = 0.52\textwidth]{TOC.pdf}
%\end{wrapfigure}

%\begin{figure}
%\includegraphics[width = \textwidth]{TOC}
%\end{figure}

%Some journals require a graphical entry for the Table of Contents.
%This should be laid out ``print ready'' so that the sizing of the

%Inside the \texttt{tocentry} environment, the font used is Helvetica
%8\,pt, as required by \emph{Journal of the American Chemical

%The surrounding frame is 9\,cm by 3.5\,cm, which is the maximum
%permitted for  \emph{Journal of the American Chemical Society}
%graphical table of content entries. The box will not resize if the
%content is too big: instead it will overflow the edge of the box.

%This box and the associated title will always be printed on a
%separate page at the end of the document.

%\end{tocentry}

\begin{abstract}

The production of high-quality two-dimensional (2D) materials is essential for the ultimate performance of single layers and their hybrids. Hexagonal boron nitride ($h$-BN) is foreseen to become the key 2D hybrid and packaging material since it is insulating, tight, flat, transparent and chemically inert, though it is difficult to attain in ultimate quality.   
Here, a new scheme is reported for producing single layer $h$-BN that shows higher quality and much more uniformity than material from chemical vapor deposition (CVD). We delaminate CVD $h$-BN from Rh(111) and transfer it to a clean metal surface. The twisting angle between BN and the new substrate yields metastable moir\'e structures. Annealing above 1000 K leads to 2D distillation, i.e., catalyst-assisted BN sublimation from the edges of the transferred layer and subsequent condensation into superior quality $h$-BN. This provides a new and low-cost way of high-quality 2D material production remote from CVD instrumentation.

\end{abstract}


\section{Introduction}

The paradigm "two-dimensional (2D) materials" \cite{Geim2013} is expected to enable unprecedented opportunities for new devices with ultimately thin membranes \cite{Surwade2015}, mechanical detectors \cite{Cartamil-Bueno2017}, inks \cite{Torrisi2012} and specifically it opens perspectives for electronics beyond silicon technology \cite{Novoselov2004,Banszerus2015}.
These new materials can be assembled at room temperature, layer by layer, which allows to produce non-equilibrium hybrid structures with unprecedented properties. For instance, the twisting of two stacked graphene layers leads to 2D superconductivity that is related to the moir\'e interference between the two 2D lattices  \cite{Cao20181}.

Before harvesting the benefits of materials, their fabrication and handling have to be mastered. This is a big challenge for 2D materials since they mostly consist of surfaces, which are known to be prone to imperfections like contaminations and defects. The related reactivity of 2D materials calls for a packaging material that is protective, tight, thin, flat, transparent, and non-reactive. Hexagonal boron nitride (\mbox{$h$-BN}) has all these properties and is the prime candidate to become the key 2D packaging and hybrid material \cite{Dean2010}.
There are however no scalable methods for the production of \mbox{$h$-BN} with a quality that can rival mechanically exfoliated $h$-BN from state-of-the-art synthetic single crystals \cite{Watanabe2004}. %which is capable of producing BN with outstanding quality but only in trace amounts.
Currently the main approach for scalable production is chemical vapour deposition (CVD) of precursors that contain boron and nitrogen \cite{Nagashima1995,Corso2004,Lee2018,Chen2020}, and segregation-assisted growth \cite{Suzuki2012,Zhang2014} on substrates acting as catalysts or bulk reservoirs of B and N.  Although CVD does not allow the preparation of single orientation $h$-BN on all substrates, transfer of single-orientation $h$-BN does, with the added feature of non-equilibrium moir\'e formation with arbitrary lattice orientation.  Yet, transfer is another challenge, since it is, if successful, often accompanies with compromises to the material quality \cite{Cun2018,Hemmi2019}. 

In view of these problems and opportunities, we transferred CVD-grown single layer boron nitride back onto a crystalline catalyst. The choice of twist angle enables moir\'e interference between the new substrate and the $h$-BN layer. Annealing of such structures reveals the temperature window, within which this moir\'e structure is efficiently cleaned and stable. At higher temperatures this metastable moir\'e undergoes a phase transformation where boron nitride sublimates from the edge of the transferred BN into a dilute adsorbate phase and a re-condensation into a lattice aligned higher quality $h$-BN. Because the sublimation and re-condensation processes are well confined to the substrate surface, we label it "2D distillation". This is a new way for producing $h$-BN remote from CVD equipment. All these products and processes can directly be monitored in real time with low energy electron microscopy (LEEM).

Single orientation $h$-BN monolayers on rhodium, also called "nanomesh" \cite{Corso2004,Berner2007}, are fabricated in a CVD process on single crystalline Rh(111) at 4-inch wafer scale \cite{Gsell2009}. Like on Ru(0001)\cite{Goriachko2007} the interaction of BN with the substrate is relatively strong, and aligned $h$-BN can be grown, while e.g. no pronounced azimuthal lock-in was observed for Pd(111)\cite{Morscher2006}.

\section{Results and Discussions}

\begin{figure}[H]
\begin{center}
\includegraphics[scale=0.68]{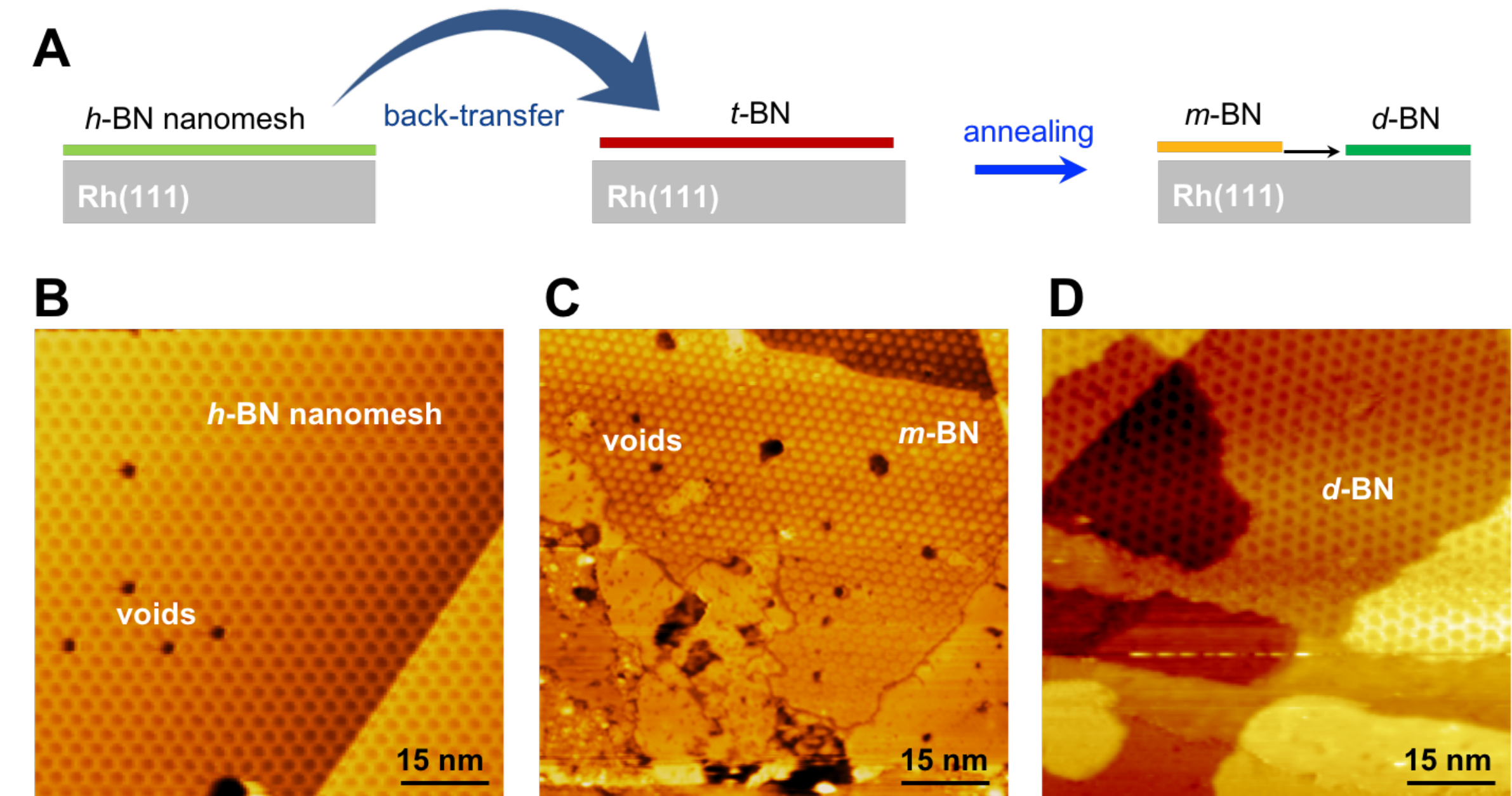}
\caption{{\textbf{\small Transfer and annealing of $h$-BN on Rh(111).}
(A) Concept: After growth and delamination, the $h$-BN monolayer is transferred on a Rh(111) substrate, where {\mbox{$t$-BN}} forms. Annealing of $t$-BN leads to visible moir\'e structures ($m$-BN) and at higher temperature $m$-BN distills into $d$-BN, i.e., nanomesh reforms. (B-D) Room temperature STM (80~$\times$~80~nm$^2$) images of pristine $h$-BN nanomesh with 2~nm voids on Rh(111) (B), $m$-BN with voids on Rh(111) after transfer and annealing to 750 K (C), and distilled $m$-BN without voids on Rh(111) ($d$-BN) after transferred and annealing above 1000 K. U = -1.20 V, I = 0.50 nA.}}
\label{F1}
\end{center}
\end{figure}

Figure \ref{F1}A depicts the procedure of the experiments. After CVD growth, a $h$-BN monolayer was transferred (named $t$-BN) on arbitrary substrates \cite{Cun2018,Hemmi2019}. Here we investigate the transfer of $h$-BN onto Rh(111). X-ray photoelectron spectroscopy (XPS) data in Figure S1 of the supporting information (SI) confirm successful transfer of $h$-BN. Notably the lattices of $t$-BN and the Rh substrate are not aligned, as it is the case for the $h$-BN nanomesh.
Annealing to 750~K cleans the surface, which induces the formation of moir\'e structures ($m$-BN) that can be resolved with scanning tunneling microscopy (STM). The $m$-BN is metastable and annealing to higher temperatures leads to %BN sublimation and re-condensation in an aligned and 
a more stable phase ($d$-BN): The process that we call "2D distillation". 
Figure \ref{F1}B shows a STM image of a CVD-grown $h$-BN nanomesh on Rh(111)  \cite{Corso2004}. Prior to the imaging, 2~nm voids have been created with the "can-opener" effect \cite{Cun2013,Cun20142} in order to have the fingerprint of the original BN layer after transfer \cite{Cun2018}.  
Figure \ref{F1}C displays a representative STM image of \mbox{$m$-BN} with such 2~nm voids. Clearly, a hexagonal super-lattice is distinguished. Compared to the nanomesh in Figure \ref{F1}B, the moir\'e lattice constant is \mbox{25\%} smaller, and the unit cell displays a 0.1~nm protrusion and not a depression (more data can be found in Figure S2 of the SI). 
Applying the theory for a periodic overlayer moir\'e pattern \cite{Hermann2012},
the super lattice constant of 2.4~nm indicates that this transferred \mbox{$t$-BN} flake was rotated by $\alpha$=4$^\circ$ with respect to the Rh lattice. 
Figure \ref{F1}D shows a region of the surface after further annealing to above 1000 K. The formation of a superstructure with a lattice constant of 3.2 nm which is within the error bar that of the $h$-BN nanomesh is observed. Furthermore, the 2 nm voids as seen in Figure \ref{F1}B\&1C disappeared. 

\begin{figure}[H]
\begin{center}
\includegraphics[scale=0.83]{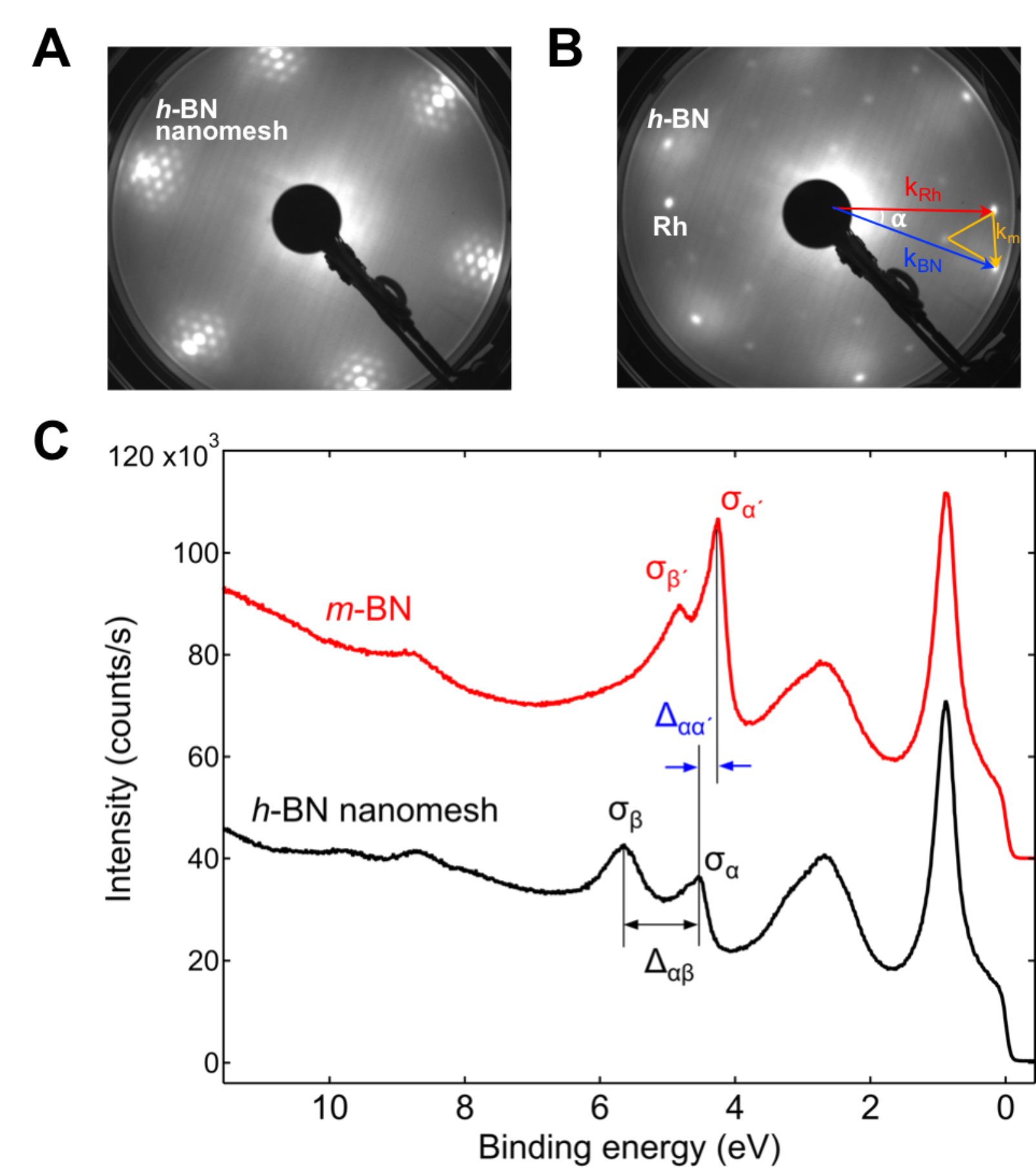}
\caption{{\textbf{\small Structural and electronic properties of $h$-BN/Rh(111) and $m$-BN/Rh(111).}
(A\&B) LEED patterns  \mbox{(E=70 eV)} of pristine $h$-BN nanomesh (A), and transferred $h$-BN on Rh(111) after annealing to 750 K ($m$-BN) (B). The reciprocal lattice vectors of Rh(111), $h$-BN and the superstructures are indicated. The angle $\alpha$ between the substrate $\vec{k}_{Rh}$ (red) and $h$-BN $\langle 10\rangle$ diffraction spots $\vec{k}_{BN}$  (blue)  is 19$^\circ$. $\vec{k}_{BN}-\vec{k}_{Rh}$ is the reciprocal moir\'e lattice vector 
$\vec{k}_{m}$ (orange). \mbox{(C) UPS (He I$_\alpha$}) spectra of pristine $h$-BN nanomesh (black) and \mbox{$m$-BN} in (B) (red). The work function increase of \mbox{$m$-BN}/Rh(111) is reflected in the $\sigma$ band shift $\Delta_{\alpha\alpha'}$. The $\sigma$-band splitting $\Delta_{\alpha\beta}$ is smaller in $m$-BN.}}
\label{F2}
\end{center}
\end{figure}

For the study of macro- and mesoscopic properties of transferred BN at the millimeter scale,  other methods than STM have to be employed. While low energy electron diffraction (LEED) maps the macroscopic crystallinity, ultraviolet photoelectron spectroscopy (UPS) provides details on the electronic structures. 
An $m$-BN sample with a lattice rotation angle~$\alpha$=19$^\circ$ and a transfer rate of 95\% is compared with a nanomesh ($\alpha$=0$^\circ$) in Figure \ref{F2}. The moir\'e is formed after transfer and annealing to 750 K. The angle $\alpha$ between the $[1\bar{1}0]$ direction of the substrate and the $[10 ]$ direction of the adsorbate is determined from the LEED patterns in \mbox{Figure \ref{F2}A\&2B}.
Normal emission angle-resolved photoemission spectroscopy (ARPES) displays a $\sigma$ band splitting $\Delta_{\alpha'\beta'}$ of 0.6 eV for this $m$-BN and an up-shift $\Delta_{\alpha\alpha'}$ of the $\sigma_\alpha$ bands by 0.27~eV to 4.27~eV binding energy.
This up-shift is in line with the $h$-BN physisorption picture \cite{Nagashima1995}, which indicates for the present case that the work function of the $\alpha$=19$^\circ$ moir\'e is 0.2~eV higher than that of the $h$-BN nanomesh.
The $\sigma_\alpha$ and $\sigma_\beta$ band-positions are measures for the electrostatic potential variations in the boron nitride super cell, and it can be seen that this splitting decreases by about a factor of two from the nanomesh to the  $\alpha$=19$^\circ$ moir\'e. The fact that the supercell size decreases by a factor of 4.24 implies even larger lateral electric fields than in the $h$-BN nanomesh \cite{Dil2008}.
This interpretation of the photoemission results is supported by density functional theory (DFT) with the calculation of 19 $h$-BN units rotated by an angle of 23.4$^\circ$ on top of a (4$\times$4) Rh unit cell. It predicts that the bonding of the rotated $h$-BN layer with the substrate is about 20\% weaker than the aligned 169 on 144 R0$^\circ$ $h$-BN/Rh(111) nanomesh (see S4 in the SI) \cite{Iannuzzi2014}.  

The properties of 2D materials strongly depend on their lattice: Strain fluctuations, and lattice rotation angle distributions at the micrometer-scale are decisive. For the case of graphene lattice strain may be measured with Raman spectroscopy \cite{Huang2009,Couto2014}. If it comes to a non-invasive method that captures the lattice rotation and the related moir\'e angles on the mesoscopic scale, LEEM gives direct insight \cite{Sutter2008,Man2012}. 
LEEM identifies different phases in bright field images, where the specular electron reflectivity is measured, while the local crystal lattice orientation and straining can be inferred from $\mu$-LEED diffraction patterns measured on 250~nm length scale \cite{Man2011}. 
Furthermore, LEEM can be performed at high temperatures, which allows to directly observe phase transformations from $t$-BN into $d$-BN in real space and time.

\begin{figure}[H]
\begin{center}
\includegraphics[scale=0.82]{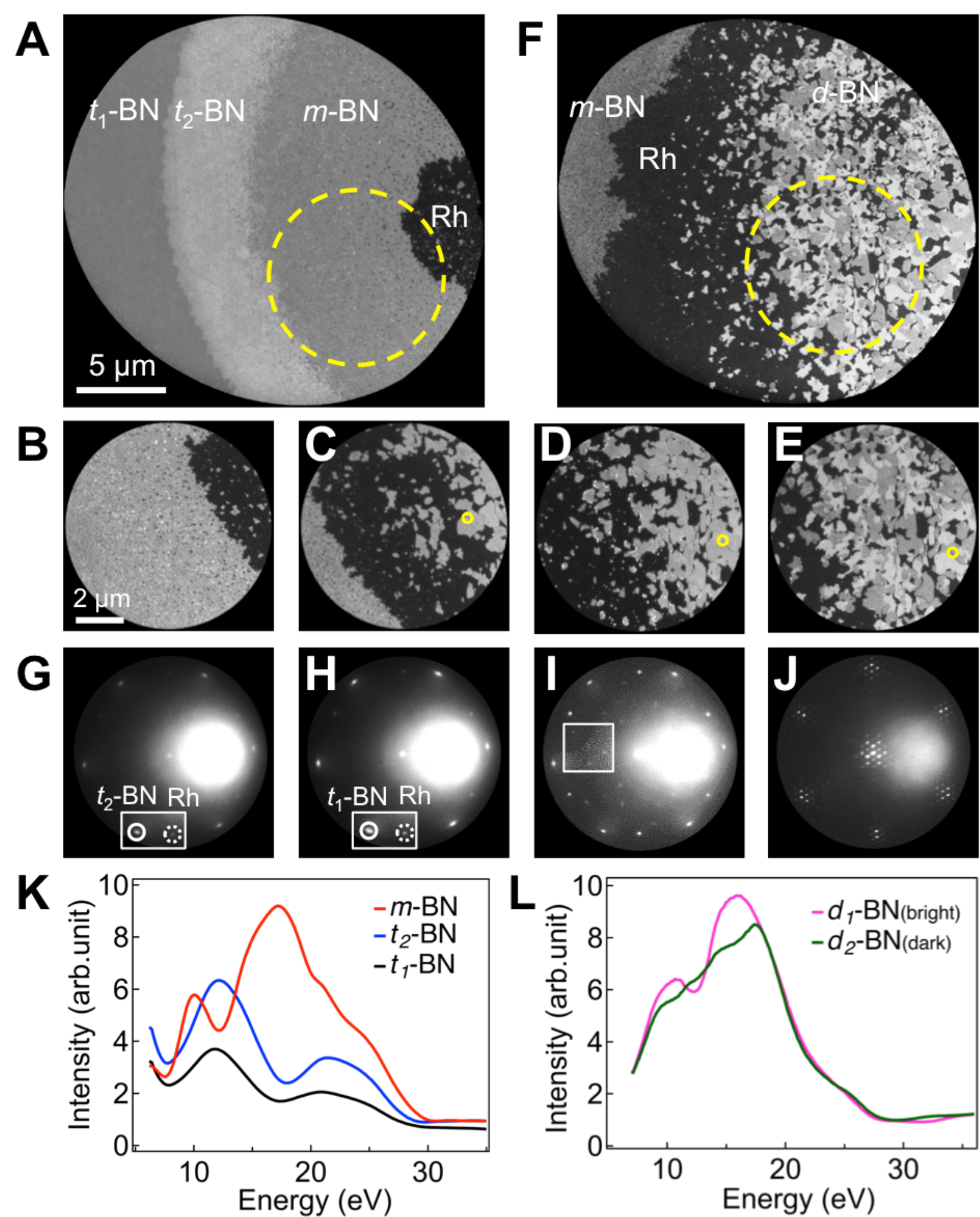}
  \caption{{\textbf{\small Low electron energy microscopy (LEEM) images, patterns and spectra.}}
(A-F) LEEM bright-field image sequence of the transformation from $t$-BN to $m$-BN, and 2D distillation to $d$-BN upon annealing (see LEEM video in the SI). The yellow circle in (C-E) is a marker at the same location on the sample to guide the viewers for how the $d$-BN evolves. (A) 1130~K, t = 0; (B) 1135~K, t = 11 min; (C) 1180~K, t = 19 min; (D) 1180~K, t = 28 min; (E) 1210~K, t = 38 min; (F) 1210~K, t = 47 min. (B-E) show zoom-ins of the area indicated by the yellow dashed-circles in (A)\&(F). The imaging energies are 12 eV in (A)-(D) and 15 eV in (E) and (F). The $\mu$-LEED patterns of (G) $t_2$-BN, (H) $t_1$-BN, (I) $m$-BN, (J) $d$-BN were recorded at room temperature after annealing to 1100~K (G-I) and 1210~K (J). $t_2$-BN, $t_1$-BN and Rh spots are highlighted by white solid-line and dashed-line circles in (G) and (H). (K,L) LEEM I(V) spectra of different phases are measured after recording the corresponding $\mu$-LEED patterns. (K) $t_1$-BN (black), $t_2$-BN (blue) and $m$-BN (red). (L) $d$-BN twin domains of $d_1$ (green) and $d_2$ (magenta).}
\label{F3}
\end{center}
\end{figure}

Figure \ref{F3} shows LEEM data from a single layer $h$-BN with 2~nm voids that was transferred on a Rh(111) thin film substrate and annealed to different temperatures.
Figure~\ref{F3}A discerns $t$-BN and $m$-BN phases within the field-of-view of 23~$\mu$m at 1130~K. The transformation reaction starts at catalytic particle sites (condensation seed, see S5.2 and Figure S7 in the SI). 
The zoom-in area (yellow dashed circle in Figure \ref{F3}A\&3F) in Figure~\ref{F3}B-3E document the 2D distillation process of $m$-BN to $d$-BN and show that the reaction proceeds with speeds in the order of 10~nm/s without the presence of BN precursor molecules in the gas phase and with less than 50\% loss of BN (see the LEEM movie and analysis details in S5.1 of the SI). 
In Figure~\ref{F3}F, the $d$-BN structure dominates, where straight segments at the edges of the $d$-BN patches reveal the crystallographic orientation of the $d$-BN.
%the registry of the $d$-BN with respect to the substrate becomes visible with the orientation of the edges of the $d$-BN patches. 
Furthermore, $d$-BN exhibits two intensities in Figure~\ref{F3}F that identify BN/substrate twinning domains.
The structural natures of the different crystalline phases are distinguished by the $\mu$-LEED patterns in Figs.~\ref{F3}G-3J obtained from 250 nm diameter areas. The rotated $t$-BN layer exhibits only Rh and rotated BN integer diffraction spots.
In the $m$-BN phase the diffraction spots of the moir\'e lattice are visible and from the lower photoelectron yield of $m$-BN we infer a lower work function of $t$-BN (details see S5.4 in the SI). Finally, the diffraction pattern of $d$-BN is reminiscent to $h$-BN/Rh(111) nanomesh with a commensurate 13$\times$13 on 12$\times$12 superstructure.
The variation of the elastically scattered intensity along the surface normal versus incident electron energy in I(V) spectra gives rise to the energy dependence of bright field LEEM image contrast and also distinguishes various phases. 
%furthermore identifies the phases in current versus voltage I(V) spectra, which explains the electron energy dependence of the bright field image contrast. 
The I(V) spectra of $t_1$-BN and $t_2$-BN in Figure~\ref{F3}A are displayed in Figure~\ref{F3}K and show similar characteristics, while $m$-BN is distinctly different. The similarity of the I(V) spectra for $t_1$-BN and $t_2$-BN indicate that they have the same structure, but the higher intensity of $t_2$-BN means that it is cleaner. This cleaning process precedes the 2D distillation (see the video in the SI).  
Figure~\ref{F3}L displays the I(V) spectra for two different BN domains $d_1$-BN and $d_2$-BN in Figure~\ref{F3}F.

The in-situ LEEM observations directly reveal the 2D distillation process, i.e., the sublimation of BN onto the Rh(111) surface, the diffusion of these precursors and condensation into a more stable material: $d$-BN. 
%Based on these in-situ electron microscopy observations, the 2D distillation picture, i.e., the 2D evaporation of BN onto the Rh(111) surface, the diffusion of these precursors and condensation into a new, more stable material like the $h$-BN nanomesh, is substantiated. 
The residence-times of B and N on the surface are sufficiently long for this process to take place. Free $h$-BN fragments continue to be generated at the receding \mbox{$m$-BN} edge and diffuse towards the $d$-BN phase in response to the free energy difference between the $m$-BN and $d$-BN edges, respectively. 

The $m$-BN $\rightarrow d$-BN phase transformation is initiated at sparsely distributed defect-sites (Figure~S7). These defects are believed to be Rh particles bound to the underside of the transferred $h$-BN that were detached from the growth substrate during the transfer process. The areal density of the particles is in the order of 10$^{-4}\mu$m$^{-2}$.
These point defects trigger the 2D distillation at low temperature, which is favorable for minimizing loss of BN from the surface \cite{Hemmi2019} and explain the circular shape of the phase transformation region. The choice of the catalyst surface plays an important role, as there is no evidence for 2D distillation in control experiments with $h$-BN on a SiO$_2$ surface (Figure S3).  

\begin{figure}[H]
\begin{center}
\includegraphics[scale=0.85]{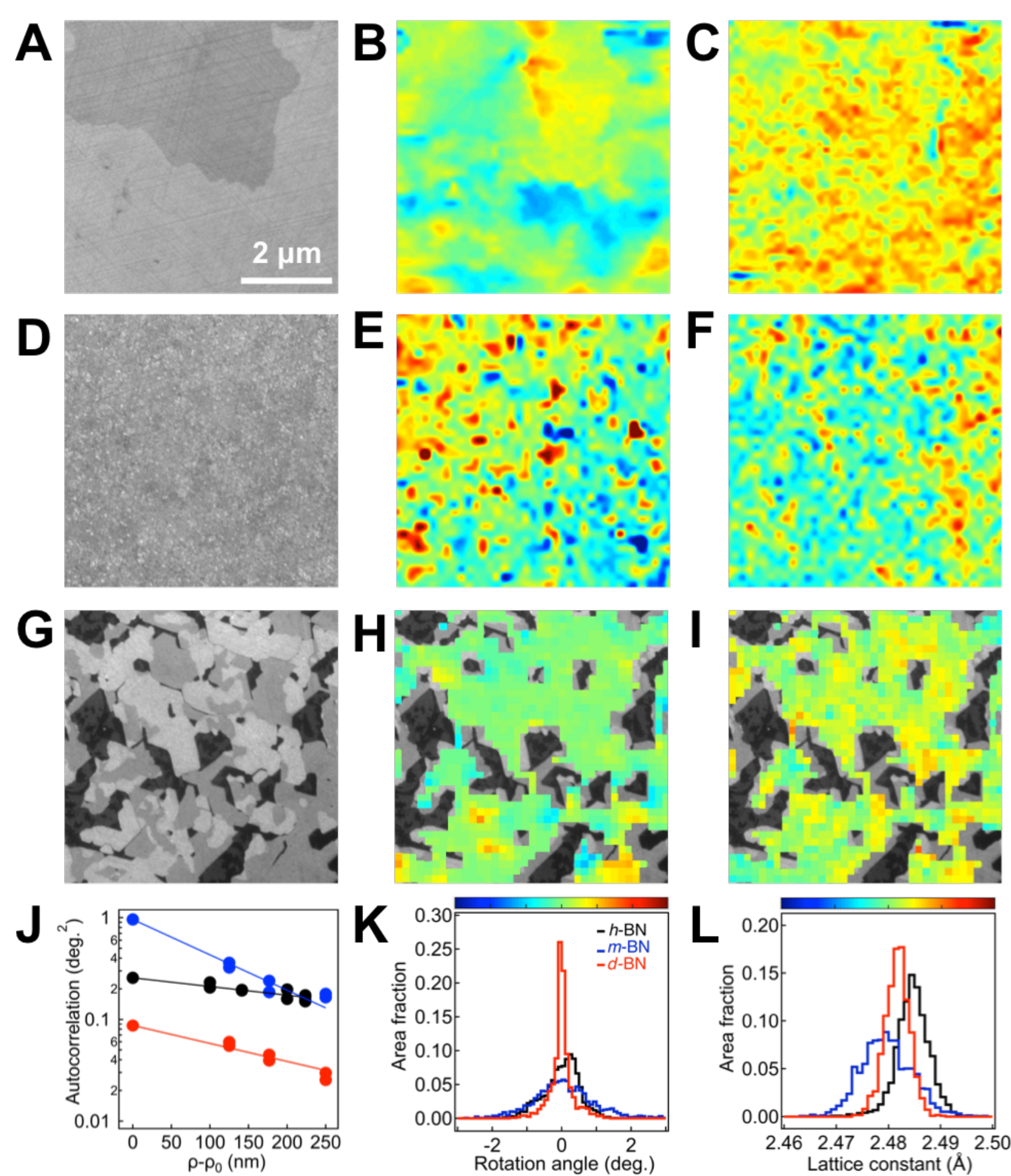}
\caption{{\textbf{\small BN lattice properties from scanning $\mu$-LEED at room temperature.}}
(A-C) $h$-BN nanomesh grown on Rh(111) and annealed to 1210 K. (D-F) $m$-BN after annealing to 1100~K. (G-I) $d$-BN distilled from $m$-BN up to 1210 K. (A,D\&G) LEEM bright field images. Color-coded (B,E\&H) lattice rotation angles, $\alpha$, and (C,F\&I) lattice constants. (J-L) Characteristics of $h$-BN (black),  $m$-BN (blue) and $d$-BN (red). (J) Spatial autocorrelation of lattice rotation angles. (K) Histogram of the lattice rotation angles. The average $m$-BN rotation angle of 24.3$^\circ$ is subtracted. (L) Histogram of the lattice constants.}
\label{F4}
\end{center}
\end{figure}

In order to establish differences in the characteristics of $h$-BN nanomesh, $m$-BN and $d$-BN, lattice constants, lattice rotation angles, and mosaicities were determined
quantitatively in submicron (250~nm) areas with high accuracy using $\mu$-LEED. These characteristics are particularly relevant for the electronic performance of 2D materials.
Figure \ref{F4}  shows $5\times 5~\mu$m$^2$ scanning {\mbox{$\mu$-LEED}} images of a $h$-BN nanomesh,  $m$-BN and $d$-BN. 
As for $d$-BN, $h$-BN may display in different domains as well, which explains the bright field contrast in Figure~\ref{F4}A, while such contrast is absent in the $m$-BN patch. 
The BN lattice rotation angles $\alpha$ and lattice constants $a$ have been determined from more than 1000 different $\mu$-LEED patterns recorded in the scan area. 
The BN lattice constants of $d$-BN have the smallest scatter of 0.1\% while for $m$-BN it is about a factor of two larger. 
The average BN lattice rotation angle of $m$-BN is 24.3$^\circ$, while $h$-BN and $d$-BN appear aligned to the Rh substrate. The mosaic spread or standard deviation of these angles is the smallest for $d$-BN. This is also reflected in the prefactors of the lateral autocorrelation of the rotation angles and we find the best correlation length $\lambda_\alpha$ of 510~nm for $h$-BN nanomesh, while it is 220~nm for $d$-BN and 130~nm for $m$-BN (see detailed analysis in S5.5 in the SI).

\section{Conclusions}

In summary, 2D distillation signifies besides gas-phase and precipitation induced growth a new way to synthesize 2D materials on surfaces. We investigated hexagonal boron nitride that is transferred back onto a rhodium substrate.
Annealing of an arbitrarily oriented transferred single layer cleans the BN layer and leads to moir\'e interference with the substrate. Further annealing treatments lead to formation of $d$-BN by sublimation and recondesation with distinctly superior quality compared with the  initially grown $h$-BN. 
%This transformation can be ignited at catalytic defects and is called 2D distillation. 
%Our results provide a facile and low-cost method for fabricating high-quality $h$-BN monolayers without CVD setup, and the general applicability of the concept of 2D distillation promises a new synthesis method for any high-quality 2D material on catalytic substrates.
Our results provide a facile and low-cost method for fabricating high-quality $h$-BN monolayers without CVD setup. The general applicability of the concept of 2D distillation promises a new synthesis method for any high-quality 2D material on catalytic substrates.

\section{Materials and Methods}
{\bf{CVD, delamination, XPS, UPS, LEED, STM \& Raman experiments in Z\"urich}}:
The \mbox{$h$-BN} monolayer is prepared on 4-inch Rh(111) thin film wafers with the procedures and facility described previously \cite{Hemmi2014}. 
The electrochemical delamination procedure of $h$-BN monolayer is described in detail previously \cite{Cun2018,Hemmi2019}.
%The electrochemical alkylation is achieved in Ar degassed, dehydrated acetonitrile and a home-build potentiostat, employing a Keithley 2602B source measure unit and a National Instruments USB-6351 card to measure the reference electrode potential. Chemicals are purchased from Fluke. The sample cleaning procedure after the TOABr/acetonitrile treatment and before any other measurement is described in detail in section 2 of SI.
XPS and UPS measurements were carried out in a VGESCALAB 220 system \cite{Greber1997}. The STM experiments were performed in a variable-temperature scanning tunneling microscope (Omicron, VT-STM) \cite{Ma2010}. 

{\bf{LEEM experiments}}: LEEM imaging and I(V) spectra measurements, conventional and scanning $\mu$-LEED and PEEM were carried out in a non-commercial LEEM instrument\cite{Altman2010,Man2011,Bauer2014,Yu2019}. PEEM measurements were performed using illumination from a mercury discharge lamp. Temperature was measured using an optical pyrometer with an emissivity setting of 0.1.

{\bf{DFT Simulations}}:
DFT simulations have been performed with the CP2K code\cite{hutter2014cp2k} using the PBE-rVV10 density functional\cite{perdew1996,vydrov2009,sabatini2013}. The ground state calculations have been carried out under the Gaussian plane wave method, the molecular orbitals of the valence electrons are expanded into a combination of Gaussian and plane waves, whereas the core electrons are treated using Goedecker-Teter-Hutter pseudopotentials \cite{goedecker1996}. 
For the Gaussian basis set expansion, the valence orbitals have been expanded into molecularly optimized DZVP basis sets for all elements except Rh, where orbitals were expanded using a molecularly optimized SZVP basis \cite{molopt}.

%\section{Extra information when writing JACS Communications}

%When producing communications for \emph{J.~Am.\ Chem.\ Soc.}, the class will automatically lay the text out in the style of the journal. This gives a guide to the length of text that can be accommodated in such a publication. There are some points to bear in mind when preparing a JACS Communication in this way.  The layout produced here is a \emph{model} for the published result, and the outcome should be taken as a \emph{guide} to the final length. The spacing and sizing of graphical content is an area where there is some flexibility in the process.  You should not worry about the space before and after graphics, which is set to give a guide to the published size. This is very dependant on the final published layout.

%You should be able to use the same source to produce a JACS Communication and a normal article.  For example, this demonstration file will work with both \texttt{type=article} and \texttt{type=communication}. Sections and any abstract are automatically ignored, although you will get warnings to this effect.

\begin{acknowledgement}

Financial support by the European Commission under the Graphene Flagship Core 2 (No. 785219) and the  HKUST Block Grant DSCI17SC02 are gratefully acknowledged. 
We thank Mr. Michael Weinl and Dr. Matthias Schreck from Universit\"{a}t Augsburg, Germany for providing Rh metal substrates.

\end{acknowledgement}

\begin{suppinfo}

%A listing of the contents of each file supplied as Supporting Information should be included. For instructions on what should be included in the Supporting Information as well as how to prepare this material for publications, refer to the journal's Instructions for Authors.

Supporting information on XPS of back-transferred $h$-BN on Rh(111), other moir\'e patterns, transferred $h$-BN on SiO$_2$, DFT calculations, LEEM movie and analysis, condensation seed, the ($2\sqrt{3}\times 2\sqrt{3})R30^\circ$ superstructure, recording sites of $\mu$-LEED and I(V) curves, and $\mu$-LEED evaluations, are available in Ref. \cite{SI}.

\end{suppinfo}

\bibliography{2D distillation_bibliography}

\newpage

\end{document}